\documentclass[12pt]{article}
\usepackage{amssymb}
\usepackage{amsfonts}
\usepackage{amsmath}
\usepackage{a4wide}
\usepackage{graphicx}

\newtheorem{proposition}{Proposition}
\newtheorem{remark}{Remark}

\newenvironment{proof}[1][Proof]{\textbf{#1.} }{\ \rule{0.5em}{0.5em}}
\makeatletter
\def\@removefromreset#1#2{\let\@tempb\@elt
     \def\@tempa#1{@&#1}\expandafter\let\csname @*#1*\endcsname\@tempa
     \def\@elt##1{\expandafter\ifx\csname @*##1*\endcsname\@tempa\else
    \noexpand\@elt{##1}\fi}     \expandafter\edef\csname cl@#2\endcsname{\csname cl@#2\endcsname}     \let\@elt\@tempb
     \expandafter\let\csname @*#1*\endcsname\@undefined}

\@removefromreset{equation}{section}

\@removefromreset{theorem}{section}
\makeatother
\input{tcilatex}

\begin{document}

\title{On validity of the original Bell inequality for the Werner
nonseparable state}
\author{Elena R. Loubenets \\
Applied Mathematics Department, \\
Moscow State Institute of Electronics and Mathematics}
\maketitle

\begin{abstract}
In quant-ph/0406139, we have introduced in a very general setting the new
class of quantum states, \textit{density source-operator} \textit{states},
satisfying any classical CHSH-form inequality, and shown that any separable
state belongs to this class. In the present paper, we prove that, the Werner
nonseparable state on $\mathbb{C}^{d}\otimes \mathbb{C}^{d},$ $d\geq 2,$
also belongs to the class of density source-operator states. Moreover, for
any $d\geq 3,$ the Werner state is in such a subclass of this class where
each density source-operator state satisfies also the perfect correlation
form of the original Bell inequality regardless of whether or not the Bell
perfect correlation restriction is fulfilled.
\end{abstract}

\section{Introduction}

The present paper is a sequel to [1] where, in a very general setting, we
have introduced the new class of quantum states satisfying the original CHSH
inequality [2] and, more generally, every \emph{classical}\footnote{%
The term \textquotedblright classical\textquotedblright\ specifies the
validity in the frame of classical probability.} CHSH-form inequality.

Any state in this new class admits a special type dilation to an extended
tensor product Hilbert space. We call this dilation a \emph{density
source-operator }and refer to a quantum state with this type of dilation as
a \emph{density source-operator state}. As we specify this in [1], the
existence for a quantum state of a density source-operator is a purely
geometrical property, and cannot be, in general, linked with the existence
for this state of a local hidden-variable (\textit{LHV}) model.

To analyze the relation between different classes of quantum states
satisfying a \textit{classical} CHSH-form inequality, we show in [1] that
all separable states belong to the class of density source-operator states
as a particular subclass and that, for the two qubit system, the Werner
nonseparable state [3] is a density source-operator state. It is, however,
well known [3,4] that the Werner nonseparable state on $\mathbb{C}%
^{d}\otimes \mathbb{C}^{d}$ satisfies the original CHSH inequality for any
dimension $d\geq 2.$

In the present paper, we prove:

\begin{itemize}
\item for any dimension $d\geq 2,$ the Werner nonseparable state belongs to
the class of density source-operator states on $\mathbb{C}^{d}\otimes 
\mathbb{C}^{d}$;

\item for any dimension $d\geq 3,$ the Werner nonseparable state satisfies
the original Bell inequality [5,6], in its perfect correlation form.
\end{itemize}

The latter earlier unknown property of the Werner state can be verified
experimentally.

\section{The Werner state as a density source-operator state}

Consider on $\mathbb{C}^{d}\otimes \mathbb{C}^{d},$ $\forall d\geq 2,$ the
Werner nonseparable quantum state [3]: 
\begin{eqnarray}
\rho _{d}^{(w)} &=&\frac{1}{d^{3}}I_{\mathbb{C}^{d}\otimes \mathbb{C}^{d}}+%
\frac{2}{d^{2}}P_{d}^{(-)}  \label{1} \\
&=&\frac{d+1}{d^{3}}I_{\mathbb{C}^{d}\otimes \mathbb{C}^{d}}-\frac{1}{d^{2}}%
V_{d},  \notag
\end{eqnarray}%
where 
\begin{equation}
P_{d}^{(-)}=\frac{1}{2}(I_{\mathbb{C}^{d}\otimes \mathbb{C}^{d}}-V_{d})
\label{pr}
\end{equation}%
is the orthogonal projection onto the antisymmetric subspace of $\mathbb{C}%
^{d}\otimes \mathbb{C}^{d}$ and $V_{d}$ is the flip operator on $\mathbb{C}%
^{d}\otimes \mathbb{C}^{d},$ defined by the relation:%
\begin{equation}
V_{d}(\psi _{1}\otimes \psi _{2}):=\psi _{2}\otimes \psi _{1},\text{ \ \ \ }%
\forall \psi _{1,}\text{ }\psi _{2}\in \mathbb{C}^{d}.
\end{equation}%
Recall that $V_{d}$ is self-adjoint, with 
\begin{equation}
\mathrm{tr}[V_{d}]=d,\text{ \ \ \ }(V_{d})^{2}=I_{\mathbb{C}^{d}\otimes 
\mathbb{C}^{d}}.
\end{equation}

Below, we use the notion of a density source-operator, introduced, in
general, in [1], section 3, and the notation for a density source-operator
on $\mathbb{C}^{d}\otimes \mathbb{C}^{d}\otimes \mathbb{C}^{d}$, specified
in [1], section 4.1.

\begin{proposition}
For any $d\geq 2,$ the Werner nonseparable state $\rho _{d}^{(w)}$\ belongs
to the class of density source-operator (DSO) states on $\mathbb{C}%
^{d}\otimes \mathbb{C}^{d}$.
\end{proposition}

\begin{proof}
Consider first the case $d=2.$ For the two-qubit Werner state 
\begin{equation}
\rho _{2}^{(w)}=\frac{3}{8}I_{\mathbb{C}^{2}\otimes \mathbb{C}^{2}}-\frac{1}{%
4}V_{2},
\end{equation}%
the density source-operator $T_{\blacktriangleright }^{(2)}$ on $\mathbb{C}%
^{2}\otimes \mathbb{C}^{2}\otimes \mathbb{C}^{2}$, with the partial traces 
\begin{equation}
\mathrm{tr}_{\mathbb{C}^{2}}^{(2)}[T_{\blacktriangleright }^{(2)}]=\mathrm{tr%
}_{\mathbb{C}^{2}}^{(3)}[T_{\blacktriangleright }^{(2)}]=\rho _{2}^{(w)}
\end{equation}%
over the elements standing on the $j$-th place, $j=2,3,$ of tensor products,
has been introduced in [1], section 7, example 1, and has the form: 
\begin{equation}
T_{\blacktriangleright }^{(2)}=\frac{1}{4}I_{\mathbb{C}^{2}\otimes \mathbb{C}%
^{2}\otimes \mathbb{C}^{2}}-\frac{1}{8}V_{2}\otimes I_{\mathbb{C}^{2}}-\frac{%
1}{8}(I_{\mathbb{C}^{2}}\otimes V_{2})(V_{2}\otimes I_{\mathbb{C}^{2}})(I_{%
\mathbb{C}^{2}}\otimes V_{2}).
\end{equation}%
To prove the statement for any $d\geq 3$, let us introduce on $\mathbb{C}%
^{d}\otimes \mathbb{C}^{d}\otimes \mathbb{C}^{d}$ the orthogonal projection 
\begin{eqnarray}
Q_{d} &=&\frac{1}{6}\{\text{ }I_{\mathbb{C}^{d}\otimes \mathbb{C}^{d}\otimes 
\mathbb{C}^{d}}-V_{d}\otimes I_{\mathbb{C}^{d}}-I_{\mathbb{C}^{d}}\otimes
V_{d}  \label{2} \\
&&-(I_{\mathbb{C}^{d}}\otimes V_{d})(V_{d}\otimes I_{\mathbb{C}^{d}})(I_{%
\mathbb{C}^{d}}\otimes V_{d})  \notag \\
&&+(I_{\mathbb{C}^{d}}\otimes V_{d})(V_{d}\otimes I_{\mathbb{C}%
^{d}})+(V_{d}\otimes I_{\mathbb{C}^{d}})(I_{\mathbb{C}^{d}}\otimes V_{d})%
\text{ }\},  \notag
\end{eqnarray}%
with%
\begin{equation}
\mathrm{tr}[Q_{d}]=\frac{d(d-1)(d-2)}{6}.
\end{equation}%
Since in an orthonormal basis $\{e_{n};$ $n=1,2,...,d\}$ in $\mathbb{C}^{d}$
the flip operator $V_{d}$ admits the representation 
\begin{equation}
V_{d}=\tsum_{n,m=1}^{d}|e_{n}\rangle \langle e_{m}|\otimes |e_{m}\rangle
\langle e_{n}|,  \label{12}
\end{equation}%
the projection (\ref{2}) can be written otherwise as: 
\begin{eqnarray}
6Q_{d} &=&I_{\mathbb{C}^{d}\otimes \mathbb{C}^{d}\otimes \mathbb{C}%
^{d}}-\tsum_{n,m=1}^{d}|e_{n}\rangle \langle e_{m}|\otimes |e_{m}\rangle
\langle e_{n}|\otimes I_{\mathbb{C}^{d}}  \label{gf1} \\
&&-\tsum_{n,m=1}^{d}I_{\mathbb{C}^{d}}\otimes |e_{n}\rangle \langle
e_{m}|\otimes |e_{m}\rangle \langle e_{n}|\text{ }-\tsum_{n,m=1}^{d}|e_{n}%
\rangle \langle e_{m}|\otimes I_{\mathbb{C}^{d}}\otimes |e_{m}\rangle
\langle e_{n}|\text{ }  \notag \\
&&+\tsum_{n,m,k=1}^{d}|e_{n}\rangle \langle e_{m}|\otimes |e_{m}\rangle
\langle e_{k}|\otimes |e_{k}\rangle \langle e_{n}|\text{ }%
+\tsum_{n,m,k=1}^{d}|e_{m}\rangle \langle e_{n}|\otimes |e_{k}\rangle
\langle e_{m}|\otimes |e_{n}\rangle \langle e_{k}|.  \notag
\end{eqnarray}%
From (\ref{gf1}) and (\ref{pr}) it follows that, for any $j=1,2,3,$%
\begin{eqnarray}
\mathrm{tr}_{\mathbb{C}^{d}}^{(j)}[Q_{d}] &=&\frac{d-2}{6}(I_{\mathbb{C}%
^{d}\otimes \mathbb{C}^{d}}-V_{d}) \\
&=&\frac{d-2}{3}P_{d}^{(-)}.  \notag
\end{eqnarray}%
For any dimension $d\geq 3,$ consider on $\mathbb{C}^{d}\otimes \mathbb{C}%
^{d}\otimes \mathbb{C}^{d}$ the density operator 
\begin{equation}
T_{\blacktriangleleft \blacktriangleright }^{(d)}=\frac{1}{d^{4}}I_{\mathbb{C%
}^{d}\otimes \mathbb{C}^{d}\otimes \mathbb{C}^{d}}+\frac{6}{d^{2}(d-2)}Q_{d}.
\end{equation}%
It is easy to verify that each density operator 
\begin{equation}
\mathrm{tr}_{\mathbb{C}^{d}}^{(j)}[T_{\blacktriangleleft \blacktriangleright
}^{(d)}],\text{ \ \ }\forall j=1,2,3,
\end{equation}%
on $\mathbb{C}^{d}\otimes \mathbb{C}^{d}$, reduced from $T_{%
\blacktriangleleft \blacktriangleright }^{(d)},$ coincide with the Werner
state, that is:%
\begin{equation}
\mathrm{tr}_{\mathbb{C}^{d}}^{(1)}[T_{\blacktriangleleft \blacktriangleright
}^{(d)}]=\mathrm{tr}_{\mathbb{C}^{d}}^{(2)}[T_{\blacktriangleleft
\blacktriangleright }^{(d)}]=\mathrm{tr}_{\mathbb{C}^{d}}^{(3)}[T_{%
\blacktriangleleft \blacktriangleright }^{(d)}]=\rho _{d}^{(w)}.
\label{symet}
\end{equation}%
Hence, by definition of a density source-operator for a quantum state (cf.
[1], section 3), the operator $T_{\blacktriangleleft \blacktriangleright
}^{(d)}$ represents a density source-operator for the Werner state $\rho
_{d}^{(w)}$, $\forall d\geq 3.$\newline
Thus, for any dimension $d\geq 2$, the Werner state $\rho _{d}^{(w)}$ on $%
\mathbb{C}^{d}\otimes \mathbb{C}^{d}$ is a density source-operator state.
\end{proof}

As we prove in [1], any density-source operator state satisfies every 
\textit{classical} CHSH-form inequality.

\section{The Werner state and the original Bell inequality}

Consider an Alice/Bob joint generalized quantum measurement\footnote{%
See [7,1] for the description of an Alice/Bob joint quantum measurement in a
very general setting.}, with outcomes $\left\vert \lambda _{1}\right\vert
\leq 1$ on the side of Alice and outcomes $\left\vert \lambda
_{2}\right\vert \leq 1$ on the side of Bob, performed on a bipartite quantum
system in a state $\rho $ on $\mathbb{C}^{d}\otimes \mathbb{C}^{d}$ and
described by the POV measure 
\begin{equation}
M_{1}^{(a)}(d\lambda _{1})\otimes M_{2}^{(b)}(d\lambda _{2}),  \label{io}
\end{equation}%
which is specified by a pair $(a,b)$ of some parameters on the sides of
Alice and Bob, respectively. Here, $M_{1}^{(a)}$ and $M_{2}^{(b)}$ are the
POV measures, describing, respectively, marginal experiments of Alice and
Bob.

Under the joint quantum measurement (\ref{io}), the expectation value of the
product $\lambda _{1}\lambda _{2}$ of outcomes is given by 
\begin{eqnarray}
\left\langle \lambda _{1}\lambda _{2}\right\rangle _{\rho }^{(a,b)} &:&=\int
\lambda _{1}\lambda _{2}\mathrm{tr}[\rho (M_{1}^{(a)}(d\lambda _{1})\otimes
M_{2}^{(b)}(d\lambda _{2}))] \\
&=&\mathrm{tr}[\rho (W_{1}^{(a)}\otimes W_{2}^{(b)})],  \notag
\end{eqnarray}%
with $W_{1}^{(a)},$ $W_{2}^{(b)}$ being quantum observables on $\mathbb{C}%
^{d}$, defined by the relations 
\begin{eqnarray}
W_{1}^{(a)} &:&=\int \lambda _{1}M_{1}^{(a)}(d\lambda _{1}),\text{ \ \ \ }%
||W_{1}^{(a)}||\leq 1, \\
W_{2}^{(b)} &:&=\int \lambda _{2}M_{2}^{(b)}(d\lambda _{2}),\text{ \ \ \ }%
||W_{2}^{(b)}||\leq 1,  \notag
\end{eqnarray}%
and representing the marginal experiments on the sides of Alice and Bob,
respectively.

Consider now three Alice/Bob joint quantum measurements (\ref{io}),
specified by pairs of parameters: 
\begin{equation}
(a,b_{1}),\text{ \ \ }(a,b_{2}),\text{ \ \ }(b_{1},b_{2}),
\end{equation}%
and satisfying the condition 
\begin{equation}
W_{1}^{(b_{1})}=W_{2}^{(b_{1})}.  \label{cond}
\end{equation}

As we discuss this in [7,1], the latter condition does not imply the
validity of the Bell perfect correlation/anticorrelation\footnote{%
See [5,6].} restrictions $\left\langle \lambda _{1}\lambda _{2}\right\rangle
_{\rho }^{(b_{1},b_{1})}=\pm 1$ and is usually fulfilled under Alice/Bob
joint quantum measurements. The condition (\ref{cond}) is, for example
always true under Alice/Bob projective spin measurements.

From the relation (\ref{symet}) and proposition 2 in [1], it follows that if
the above specified Alice/Bob joint measurements are performed on a
bipartite quantum system in the Werner nonseparable state $\rho _{d}^{(w)}$
on $\mathbb{C}^{d}\otimes \mathbb{C}^{d}$ then, for any dimension $d\geq 3$,
the expectation values 
\begin{equation}
\left\langle \lambda _{1}\lambda _{2}\right\rangle _{\rho
_{d}^{(w)}}^{(a,b_{1})},\text{ \ \ \ }\left\langle \lambda _{1}\lambda
_{2}\right\rangle _{\rho _{d}^{(w)}}^{(a,b_{2})},\text{ \ \ \ }\left\langle
\lambda _{1}\lambda _{2}\right\rangle _{\rho _{d}^{(w)}}^{(b_{1},b_{2})}
\end{equation}%
satisfy the perfect correlation form of the original Bell inequality, that
is: 
\begin{eqnarray}
&&\left\vert \left\langle \lambda _{1}\lambda _{2}\right\rangle _{\rho
_{d}^{(w)}}^{(a,b_{1})}-\left\langle \lambda _{1}\lambda _{2}\right\rangle
_{\rho _{d}^{(w)}}^{(a,b_{2})}\right\vert  \label{bell} \\
&=&\left\vert \text{ }\mathrm{tr}[\rho _{d}^{(w)}(W_{1}^{(a)}\otimes
W_{2}^{(b_{1})})]-\mathrm{tr}[\rho _{d}^{(w)}(W_{1}^{(a)}\otimes
W_{2}^{(b_{2})})]\text{ }\right\vert  \notag \\
&\leq &1-\mathrm{tr}[\rho _{d}^{(w)}(W_{1}^{(b_{1})}\otimes W_{2}^{(b_{2})})]
\notag \\
&=&1-\left\langle \lambda _{1}\lambda _{2}\right\rangle _{\rho
_{d}^{(w)}}^{(b_{1},b_{2})}.  \notag
\end{eqnarray}

Written otherwise, the original Bell inequality (\ref{bell}) for the Werner
state $\rho _{d}^{(w)},$ $\forall d\geq 3,$ reads 
\begin{eqnarray}
&&\left\vert \text{ }\mathrm{tr}[\rho _{d}^{(w)}(J^{(a)}\otimes J^{(b_{1})}]-%
\mathrm{tr}[\rho _{d}^{(w)}(J^{(a)}\otimes J^{(b_{2})}]\text{ }\right\vert \\
&\leq &1-\mathrm{tr}[\rho _{d}^{(w)}(J^{(b_{1})}\otimes J^{(b_{2})}]  \notag
\end{eqnarray}%
and is valid for any three quantum observables $J^{(a)},$ $J^{(b_{1})}$, $%
J^{(b_{2})}$ on $\mathbb{C}^{d},$ with the operator norms $||\cdot ||\leq 1.$

\begin{remark}
Notice that, as is the case for any density source-operator state specified
in proposition 2 in [1], for the Werner state $\rho _{d}^{(w)}$, $\forall
d\geq 3,$ the perfect correlation form of the original Bell inequality is
valid regardless of whether or not the Bell perfect correlation restriction $%
\left\langle \lambda _{1}\lambda _{2}\right\rangle _{\rho
_{d}^{(w)}}^{(b_{1},b_{1})}=1$ is satisfied.
\end{remark}

Thus, for any dimension $d\geq 3,$ the Werner nonseparable state on $\mathbb{%
C}^{d}\otimes \mathbb{C}^{d}$ satisfies both classical probability
constraints - the original CHSH inequality and the original Bell inequality,
in its perfect correlation form.

\end{document}